\documentclass[journal=nalefd,manuscript=letter,layout=twocolumn]{achemso}
\usepackage{amsmath,amssymb,amsfonts}
\usepackage[osf]{newpxtext}
\usepackage{newpxmath}
\usepackage[utf8]{inputenx}
\usepackage{graphicx}
\usepackage[dvipsnames]{xcolor}
\usepackage[pdftex]{hyperref}

\DeclareMathAlphabet{\mathcal}{OMS}{cmsy}{m}{n}

\hypersetup{
	pdfauthor={Chiara Sinito},
	bookmarksnumbered=true,
	pdftitle={Absence of quantum-confined Stark effect in GaN quantum disks embedded in (Al,Ga)N nanowires grown by molecular beam epitaxy},
	colorlinks,
	citecolor=blue,
	linkcolor=blue,
	urlcolor=blue}

\usepackage{xspace}
\usepackage{soul}

\title{Absence of quantum-confined Stark effect in GaN quantum disks embedded in (Al,Ga)N nanowires grown by molecular beam epitaxy}

\author{C.~Sinito}
\affiliation{Paul-Drude-Institut für Festkörperelektronik, Leibniz-Institut im Forschungsverbund Berlin~e.\,V., Hausvogteiplatz 5--7, 10117 Berlin, Germany}
\alsoaffiliation{Present address: Attolight AG, EPFL Innovation Park, Bât.\ D, 1015 Lausanne, Switzerland}
\email{chiara.sinito@attolight.com}
\author{P.~Corfdir}
\affiliation{Paul-Drude-Institut für Festkörperelektronik, Leibniz-Institut im Forschungsverbund Berlin~e.\,V., Hausvogteiplatz 5--7, 10117 Berlin, Germany}
\alsoaffiliation{Present address: ABB Corporate Research, 5405 Baden-Dättwil, Switzerland}
\author{C.~Pfüller}
\affiliation{Paul-Drude-Institut für Festkörperelektronik, Leibniz-Institut im Forschungsverbund Berlin~e.\,V., Hausvogteiplatz 5--7, 10117 Berlin, Germany}
\author{G.~Gao}
\affiliation{Paul-Drude-Institut für Festkörperelektronik, Leibniz-Institut im Forschungsverbund Berlin~e.\,V., Hausvogteiplatz 5--7, 10117 Berlin, Germany}
\alsoaffiliation{Present address: Materials Science and Nanoengineering Department, Rice University, Houston, Texas 77005, United States of America}
\author{J.~Bartolomé Vílchez}
\affiliation{Paul-Drude-Institut für Festkörperelektronik, Leibniz-Institut im Forschungsverbund Berlin~e.\,V., Hausvogteiplatz 5--7, 10117 Berlin, Germany}
\alsoaffiliation{Present address: Departamento de Física de Materiales, Universidad Complutense de Madrid, Madrid 28040, Spain}
\author{S.~Kölling}
\author{A.~Rodil Doblado}
\affiliation{Department of Applied Physics, TU Eindhoven, Den Dolech 2, 5612 AZ Eindhoven, The Netherlands}
\alsoaffiliation{Present address: Biomedical Neuroengineering Research Group (nBio), Bioengineering Institute, Universidad Miguel Hernández, Elche 03202, Spain}
\author{U.~Jahn}
\author{J.~Lähnemann}
\author{T.~Auzelle}
\author{J.~K.~Zettler}
\affiliation{Paul-Drude-Institut für Festkörperelektronik, Leibniz-Institut im Forschungsverbund Berlin~e.\,V., Hausvogteiplatz 5--7, 10117 Berlin, Germany}
\alsoaffiliation{Present address: LayTec AG, Seesener Str.\ 10--13, 10709 Berlin, Germany}
\author{T.~Flissikowski}
\affiliation{Paul-Drude-Institut für Festkörperelektronik, Leibniz-Institut im Forschungsverbund Berlin~e.\,V., Hausvogteiplatz 5--7, 10117 Berlin, Germany}
\author{P.~Koenraad}
\affiliation{Department of Applied Physics, TU Eindhoven, Den Dolech 2, 5612 AZ Eindhoven, The Netherlands}
\author{H.~T.~Grahn}
\author{L.~Geelhaar}
\author{S.~Fernández-Garrido}
\affiliation{Paul-Drude-Institut für Festkörperelektronik, Leibniz-Institut im Forschungsverbund Berlin~e.\,V., Hausvogteiplatz 5--7, 10117 Berlin, Germany}
\alsoaffiliation{Present address: Grupo de electrónica y semiconductores, Dpto.\ Física Aplicada, Universidad Autónoma de Madrid, C/ Francisco Tomás y Valiente 7, 28049 Madrid, Spain}
\author{O.~Brandt}
\affiliation{Paul-Drude-Institut für Festkörperelektronik, Leibniz-Institut im Forschungsverbund Berlin~e.\,V., Hausvogteiplatz 5--7, 10117 Berlin, Germany}
\email{brandt@pdi-berlin.de}

\begin{document}



\begin{abstract}

\textbf{\small Several of the key issues of planar (Al,Ga)N-based deep-ultraviolet light emitting diodes could potentially be overcome by utilizing nanowire heterostructures, exhibiting high structural perfection and improved light extraction. Here, we study the spontaneous emission of GaN/(Al,Ga)N nanowire ensembles grown on Si$(111)$ by plasma-assisted molecular beam epitaxy. The nanowires contain single GaN quantum disks embedded in long (Al,Ga)N nanowire segments essential for efficient light extraction. These quantum disks are found to exhibit intense emission at unexpectedly high energies, namely, significantly above the GaN bandgap, and almost independent of the disk thickness. An in-depth investigation of the actual structure and composition of the nanowires reveals a spontaneously formed Al gradient both along and across the nanowire, resulting in a complex core/shell structure with an Al deficient core and an Al rich shell with continuously varying Al content along the entire length of the (Al,Ga)N segment. This compositional change along the nanowire growth axis induces a polarization doping of the shell that results in a degenerate electron gas in the disk, thus screening the built-in electric fields. The high carrier density not only results in the unexpectedly high transition energies, but also in radiative lifetimes depending only weakly on temperature, leading to a comparatively high internal quantum efficiency of the GaN quantum disks up to room temperature.}

\end{abstract}



(Al,Ga)N is the material of choice for the fabrication of deep ultraviolet (UV) solid-state emitters with a wide range of applications in biological, medical, and communication areas \cite{Kneissl2011,Drost2014}. However, planar (Al,Ga)N-based heterostructures suffer from a number of problems, among them high densities of threading dislocations, low dopant activation and carrier injection, poor light extraction, and strong internal electrostatic fields in quantum wells. As a result, the currently existing (Al,Ga)N-based UV light emitting diodes (LEDs) still exhibit comparatively low external quantum efficiencies, and their performance degrades with decreasing wavelength \cite{Liao2011,Kneissl2011,Wang2012,Matafonova2018}.

(Al,Ga)N-based heterostructures in the form of three-dimensional nanostructures have the potential to resolve several of these issues challenging conventional planar devices. Consequently, the fabrication of such nanostructures is currently the subject of worldwide research activities, utilizing a wide range of techniques ranging from basic top-down \cite{Dong2014} and bottom-up \cite{Ren2018, Kim2018} approaches to hybrid schemes combining top-down processes and subsequent regrowth \cite{Tian2016, Coulon2018}. A particularly active field of research is the synthesis of GaN/(Al,Ga)N nanowires by molecular beam epitaxy (MBE) \cite{Min2018}. These nanowires tend to form on a variety of structurally and chemically dissimilar substrates, invariably yielding material of high structural perfection, since dislocations do not propagate along the nanowire growth axis \cite{Hersee2011}. In addition, the efficient elastic strain relaxation at the nanowire sidewalls facilitates the fabrication of GaN/(Al,Ga)N heterostructures with coherent interfaces. Moreover, the elastic strain relief leads to reduced polarization fields as compared to planar layers, which is expected to result in higher internal quantum efficiencies \cite{Rigutti2010}. Finally, the high geometric aspect ratio of nanowires offers superior light extraction \cite{Li2012}.

In most previous works utilizing MBE, the structures under investigation consisted of axial GaN/(Al,Ga)N nanowires containing an active region formed by GaN or (Al,Ga)N quantum disks clad by comparatively thin (Al,Ga)N or AlN barriers.\cite{Min2018} Ultimately, a nanowire heterostructure serving as efficient deep UV emitter should be free of GaN segments to avoid any reabsorption of the light emitted by the quantum disk. At the same time, a GaN quantum disk embedded in a pure (Al,Ga)N nanowire or a long (Al,Ga)N segment will exhibit a significantly larger internal field than the same disk in a GaN nanowire with short (Al,Ga)N segments.\cite{Leroux1998,Leroux1999} Since the radiative recombination rate of charge carriers confined in the disk critically depends on the magnitude of the internal electrostatic field,\cite{Polland1985} it is of high relevance for future applications to investigate the spontaneous emission of GaN quantum disks embedded in (Al,Ga)N nanowires.

In this Letter, we examine the carrier recombination of GaN/(Al,Ga)N nanowires synthesized by plasma-assisted MBE (PAMBE) on Si$(111)$ substrates. The nanowires contain a single GaN quantum disk embedded in an 0.8~\textmu m long (Al,Ga)N segment. The disks have a thickness ranging from 1.8 to 6.3~nm and emit intense luminescence at an energy invariably above the bandgap of bulk GaN, suggesting that the quantum-confined Stark effect in the disk is essentially absent. A detailed investigation of the composition of the (Al,Ga)N segment reveals a core/shell microstructure formed spontaneously during growth, with a continuously varying Al content both across the nanowire diameter and along the growth direction. The latter induces a high electron density in the shell via polarization doping, leaving the quantum disk in a degenerately doped state with screened internal electrostatic fields thus resulting in the high transition energies regardless of the disk thickness. A quantitative analysis of the transition energies and the temperature dependence of the radiative lifetime allows us to extract not only the mean value of the electron density in the disks, but also its distribution. Further consequences of the degenerate doping level are a short radiative lifetime and a comparatively high internal quantum efficiency up to room temperature.

\begin{figure*}
\includegraphics[width=0.9\textwidth]{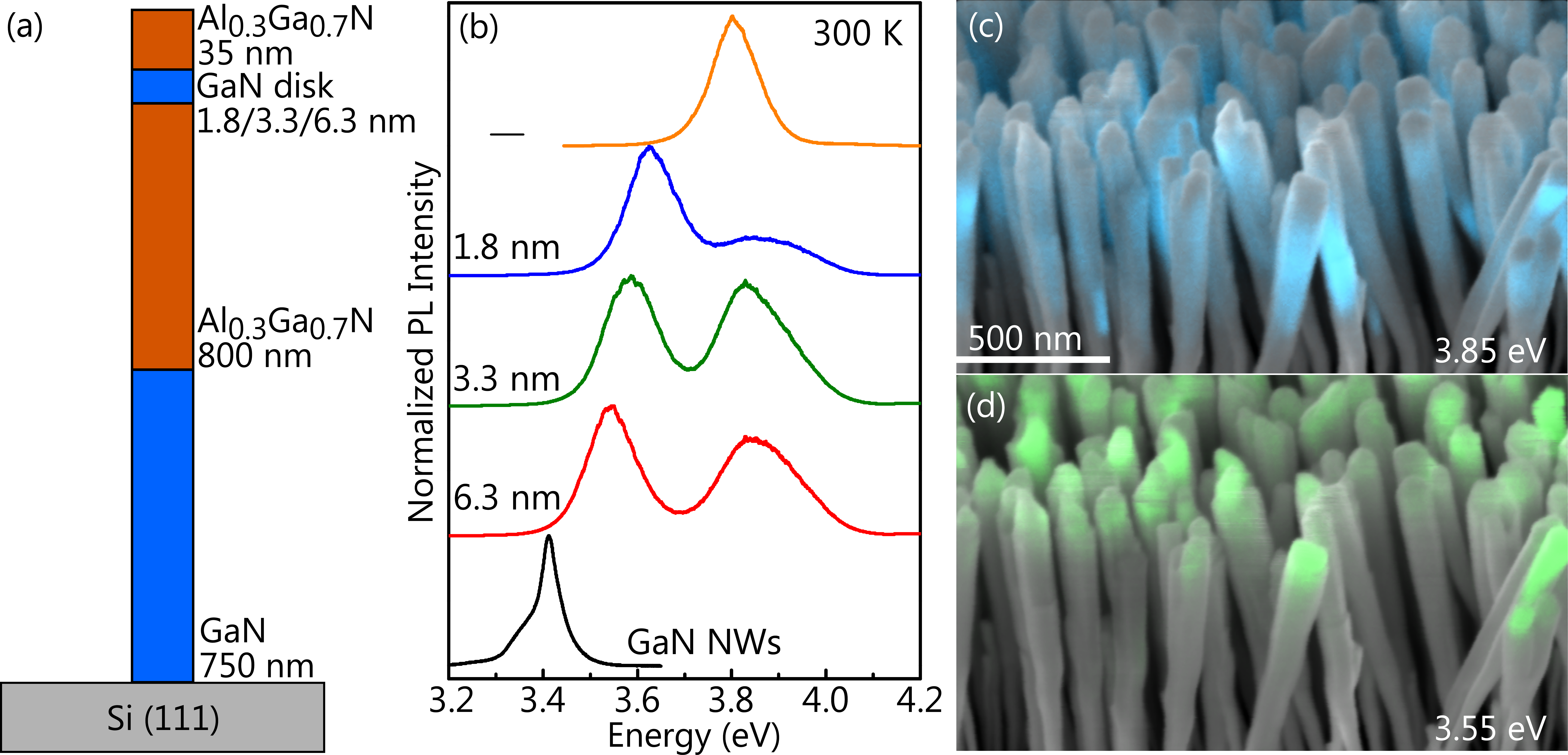}
\caption{\label{fig:PLandCL} Spectrally and spatially resolved luminescence of the GaN/(Al,Ga)N nanowires under investigation at room temperature. (a) Schematic representation of the nominal structure of the nanowires (not to scale). (b) PL spectra of the four GaN/(Al,Ga)N nanowire ensembles compared to the PL spectrum of a GaN nanowire ensemble. The spectra are normalized and vertically shifted for clarity. The numbers indicate the thickness of the GaN quantum disk for the samples containing one. (c) and (d) Monochromatic CL maps at 3.85 (c) and 3.55~eV (d) superimposed to bird's eye view secondary electron micrographs of the GaN/(Al,Ga)N nanowire ensemble with a 6.3~nm-thick GaN disk.}
\end{figure*}

Figure~\ref{fig:PLandCL}(a) shows the nominal structure of our nanowires, which consist of a 750-nm-long GaN base grown on Si$(111)$, followed by an 800-nm-long (Al,Ga)N segment and a GaN quantum disk embedded between the (Al,Ga)N segment and a 35-nm-long (Al,Ga)N cap. The samples under investigation differ in the thickness of the disk, with nominal values of 1.8, 3.3, and 6.3~nm. Both the (Al,Ga)N segment and the (Al,Ga)N cap have a nominal Al content of 0.3. For comparison, we also include a sample nominally identical to those above, but without an embedded GaN quantum disk.

Figure~\ref{fig:PLandCL}(b) depicts room temperature photoluminescence (PL) spectra of the four nanowire ensembles under continuous-wave excitation at 244~nm together with that of a bare GaN nanowire ensemble under continuous-wave excitation at 325 nm. The PL spectrum of the GaN nanowires shows a line at 3.412~eV originating from the recombination of free A excitons in strain-free GaN \cite{Hauswald2014}. The PL spectrum of the bare (Al,Ga)N nanowires exhibits a single band at 3.80~eV. A band at essentially the same spectral position (3.85~eV) is also observed in the PL spectra of the (Al,Ga)N nanowires with an embedded GaN disk, and we thus attribute this high-energy band to emission from the (Al,Ga)N segment. Consequently, the low-energy band in the spectra is associated with the GaN quantum disks. In all cases, the corresponding transition energy is notably higher than the bandgap energy of GaN and depends only weakly on the thickness of the disk itself. The same is true for the integrated PL intensity of this band. In addition, the PL intensity drops by only a factor of two to three between 10 and 300~K (see Supporting Information) and is thus about two orders of magnitude higher than that of the bare GaN nanowires at 300~K. These results suggest that the GaN quantum disks exhibit a very high internal quantum efficiency up to room temperature. As shown in the Supporting Information, the low thermal quenching is in part due to an efficient transfer of photoexcited carriers from the (Al,Ga)N segment to the quantum disk, but the actual internal quantum efficiency remains significantly higher (about a factor of 10) than that of GaN nanowires of comparable length and density.  

For the nominal parameters of our nanowires [see Fig.~\ref{fig:PLandCL}(a)], we would expect very different transition energies for both the (Al,Ga)N segment and the GaN quantum disk. Specifically, near-band edge emission of Al$_{0.3}$Ga$_{0.7}$N should occur above 4~eV,\cite{Brunner1997,Kim2000,Coli2001,Coli2002} while the energy observed (3.85~eV) is rather characteristic for an Al content of 0.15.\cite{Steude1999} Moreover, the large internal electrostatic fields in GaN/Al$_{0.3}$Ga$_{0.7}$N quantum disks should result in a quantum-confined Stark shift of typically 200--350~meV/nm,\cite{Dreyer2016} leading to a strong dependence of the transition energies on disk thickness and emission energies well below the band gap of GaN for the two thicker quantum disks.\cite{Leroux1998,Grandjean1999a,Grandjean1999b}

To confirm our assignment of the PL bands in  Fig.~\ref{fig:PLandCL}(b), we perform spatially resolved cathodoluminescence (CL) measurements at room temperature on the ensemble with the thickest GaN disk. Figures~\ref{fig:PLandCL}(c) and \ref{fig:PLandCL}(d) show monochromatic CL maps collected at emission energies of 3.85 and 3.55~eV, respectively, superimposed on the same bird's eye view secondary electron micrograph of the nanowire ensemble. The maps clearly show the complementary nature of the two emission bands, in that the band at 3.55~eV arises from the top of the nanowires, where the GaN disk is located, while the homogeneous emission at 3.85~eV originates from the region corresponding to the (Al,Ga)N segment.

\begin{figure*}[h!]
\includegraphics*[width=0.8\textwidth]{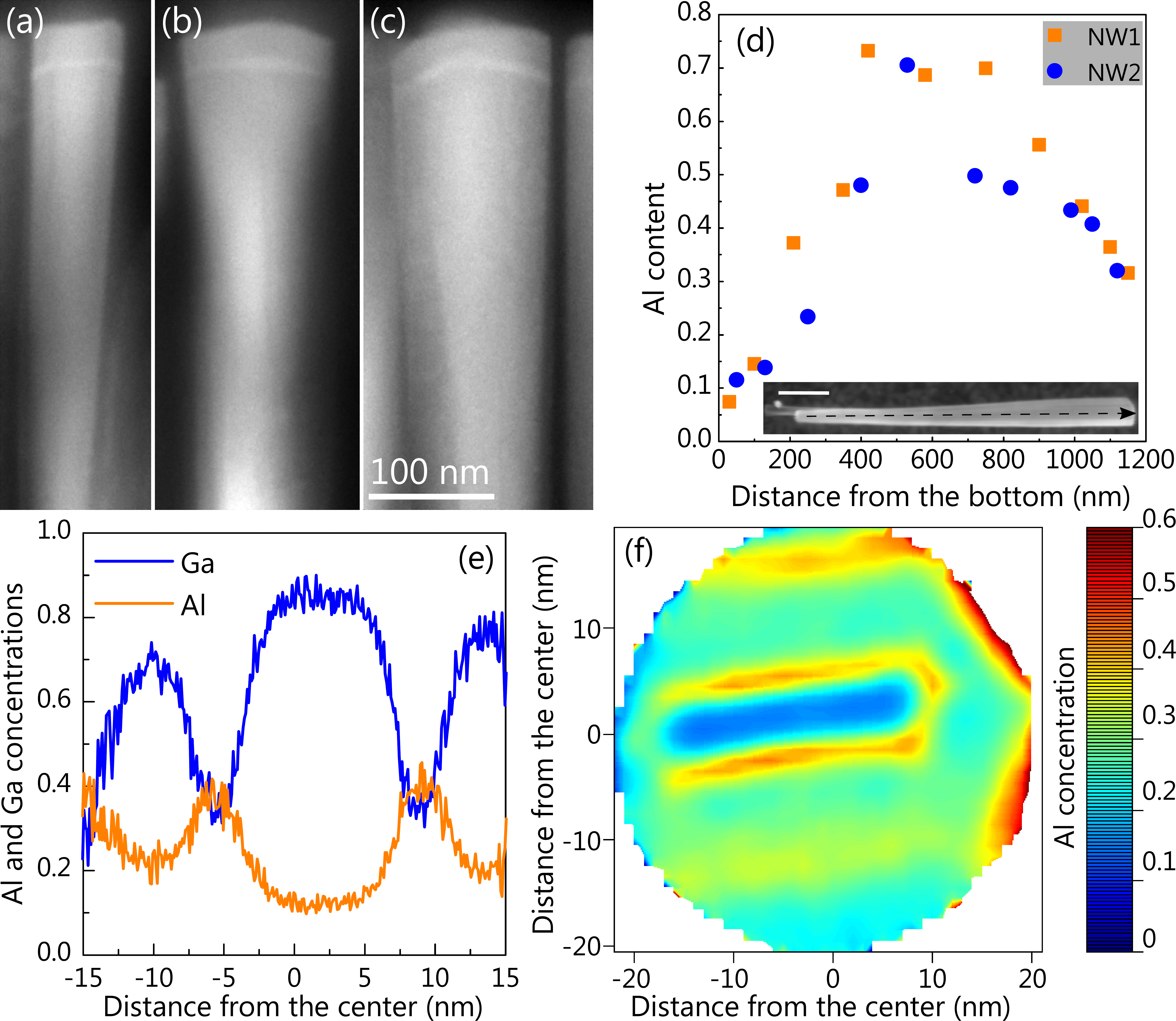}
\caption{\label{fig:structure} Structural properties of the nanowire ensemble with the 6.3-nm-thick GaN disk. (a)--(c) HAADF-STEM images of three representative nanowires. The micrographs were acquired in the $\langle 11\bar{2}0 \rangle$ zone axis. (d) Radially averaged Al content along the nanowire growth axis of two individual nanowires recorded by EDX line scans in an SEM. Inset: secondary electron micrograph of a representative nanowire with the scale bar having a length of 200~nm. Most dispersed nanowires exhibit a length between 1.0 and 1.4~\textmu m, indicating that they break off at their coalescence joints (see sideview in Supporting Information) upon ultrasonic harvesting. The arrow indicates the $[000\bar{1}]$ growth direction. (e) Ga and Al concentration profiles across the nanowire diameter in proximity of the GaN disk extracted from APT data on a single nanowire. (f) Al concentration across a two-dimensional section extracted from the same data.}
\end{figure*}

The high transition energies of the GaN quantum disks could be explained if they were actually much thinner than nominally. In this case, the quantum-confined Stark shift would be outweighed by quantum confinement, resulting in energies larger than the GaN band gap. To test this hypothesis, we perform a detailed structural and chemical analysis of the nanowire ensemble with the 6.3-nm-thick GaN disk, whose transition energy depends most sensitively on the internal electrostatic fields. Figure~\ref{fig:structure}(a)--\ref{fig:structure}(c) show scanning transmission electron microscopy (STEM) images acquired in high-angular annular dark-field (HAADF) mode for three representative single nanowires from this ensemble. The GaN quantum disk is clearly visible in all three micrographs and has a thickness of $(6 \pm 1)$~nm, confirming that the actual thickness is consistent with the nominal value. This result rules out the possibility that the high transition energies are caused by strong quantum confinement in thin disks. Moreover, the disks are seen to intersect the entire nanowire and are not enclosed in (Al,Ga)N shells, excluding also the possibility of a blueshifted emission energy due to an essentially hydrostatic strain state of the disk.

The HAADF-STEM images also reveal that the quantum disks are not delimited by flat $(000\bar{1})$ planes, but by facets with semi-polar orientation, as most clearly seen in Fig.~\ref{fig:structure}(c). The angles between the growth axis and the direction perpendicular to the semi-polar plane lie in the range of 20--30$^{\circ}$, which thus correspond to $({10\bar{1}\ell})$ planes with $\ell$ ranging from 5 to 3. The internal electrostatic field in semipolar strained GaN/(Al,Ga)N quantum wells with these orientations is reduced, but only to about 70\% of the one in a GaN/(Al,Ga)N\{$0001$\} quantum well \cite{Bigenwald2012}. Fields of this strength ($\approx 2$~MV/cm) would still lead to transition energies well below the band gap of GaN,\cite{Leroux1999} and the occurrence of semipolar facets therefore does not explain the discrepancy between the expected and experimentally observed transition energies.

What we would expect from the various shapes of the GaN quantum disks, however, is a significant variation in emission energy and thus a strong broadening of ensemble spectra such as those depicted in Fig.~\ref{fig:PLandCL}(b). Compared to the full width at half maximum (FWHM) of 40~meV for the room temperature PL line of the GaN nanowire ensemble, the PL bands of the GaN quantum disks are certainly broader, but with an FWHM of 120~meV not nearly as much as expected from the variation in thickness and orientation. The surprising uniformity of the emission energy of the GaN disks is also evident in the CL map displayed in Fig.~\ref{fig:PLandCL}(d), where the majority of the nanowires are seen to emit at 3.55~eV.

The contrast obtained by HAADF-STEM is primarily determined by the composition of the material. For several nanowires, we have observed a contrast along the nanowire cross-section as most clearly visible in Fig.~\ref{fig:structure}(a), namely, a brighter core and a darker shell, indicating the presence of a core/shell structure with an Al-deficient inner core and an Al-rich outer shell which wraps around the core. Axial contrast variations in the HAADF-STEM images are also seen in Figs.~~\ref{fig:structure}(a)--\ref{fig:structure}(c), suggesting that the Al concentration also varies along the nanowire axis. To investigate this finding in more detail, we analyze individual nanowires with chemically sensitive techniques. Figure~\ref{fig:structure}(d) shows the (radially averaged) Al content along the growth axis for two representative single nanowires as obtained by energy-dispersive x-ray spectroscopy (EDX) performed in our analytical scanning electron microscope (SEM) (see \hyperref[sec:methods]{Methods}). The inset illustrates the direction of the line scan on an exemplary nanowire. From the bottom to the top of the nanowire, the average axial Al content of the two nanowires increases from 0.1 to 0.7 up to a length of 400~nm, stays then constant for about 300~nm, and decreases from 0.7 to 0.3 for the top 500~nm of the nanowire. Given that the length of the (Al,Ga)N segment amounts to only 835~nm, the initial increase of the average Al content for the first 400~nm is likely caused by the formation of an AlN shell (with decreasing thickness toward the bottom) wrapping around the GaN base. The average Al content in proximity of the GaN disk is about 0.35, which is in fact close to the nominal value. 

The presence of this axial gradient in the radially averaged Al content cannot explain the constant, low emission energy observed along the (Al,Ga)N segment [Fig.~\ref{fig:PLandCL}(c)]. To resolve also the radial composition, we utilize atom probe tomography (APT). Figure~\ref{fig:structure}(e) shows the Ga and Al profiles across the diameter of a single nanowire in proximity of the GaN disk extracted from full three-dimensional  APT data. Several profiles have been extracted at different heights of the same nanowire (not shown), and they are all consistent with the profiles shown in Fig.~\ref{fig:structure}(e), which demonstrates the presence of an Al-deficient inner core and a complex Al-rich shell with a continuously varying Al content. The structure could also be described as Al-deficient (Al,Ga)N core wrapped by an (Al,Ga)N multi-shell structure, namely, a shell of intermediate Al content which itself is clad by thin shells with high Al content. Figure~\ref{fig:structure}(f) shows the Al content across a two-dimensional section of the same nanowire in proximity of the GaN disk. The inner blue region represents the Al-deficient core, which is seen to be surrounded by a multi-shell structure as described above. Note that the outermost thin shell with high Al content is visible only in the upper right part of the map due to the limited field-of-view and the finite tilt of the nanowire (see Supporting Information). The core appears compressed in the APT image due to well-known image aberration effects.\cite{Melkonyan2017} When correcting for the compression, the dimensions of the core would change slightly but the anisotropic shape would remain.\cite{DeGeuser2007} However, the exact shape of the core may be affected by the dependence of the evaporation rate of (Al,Ga)N on the Al content, which may potentially result in a change of the curvature of the surface of the specimen during evaporation. Note that the cation concentrations given in Figs.~\ref{fig:structure}(e) and \ref{fig:structure}(f) correspond to the raw measured data without correcting for the loss of N atoms common in APT,\cite{Riley2012} which is why the sum of the Al and Ga content does not equal one. The actual concentrations will thus be slightly higher. Considering this fact, the Al content in the core should be close to the value of 0.15 deduced from the transition energy in the PL and CL experiments shown in Fig.~\ref{fig:PLandCL}. 

\begin{figure*}[h!]
\includegraphics[width=\textwidth]{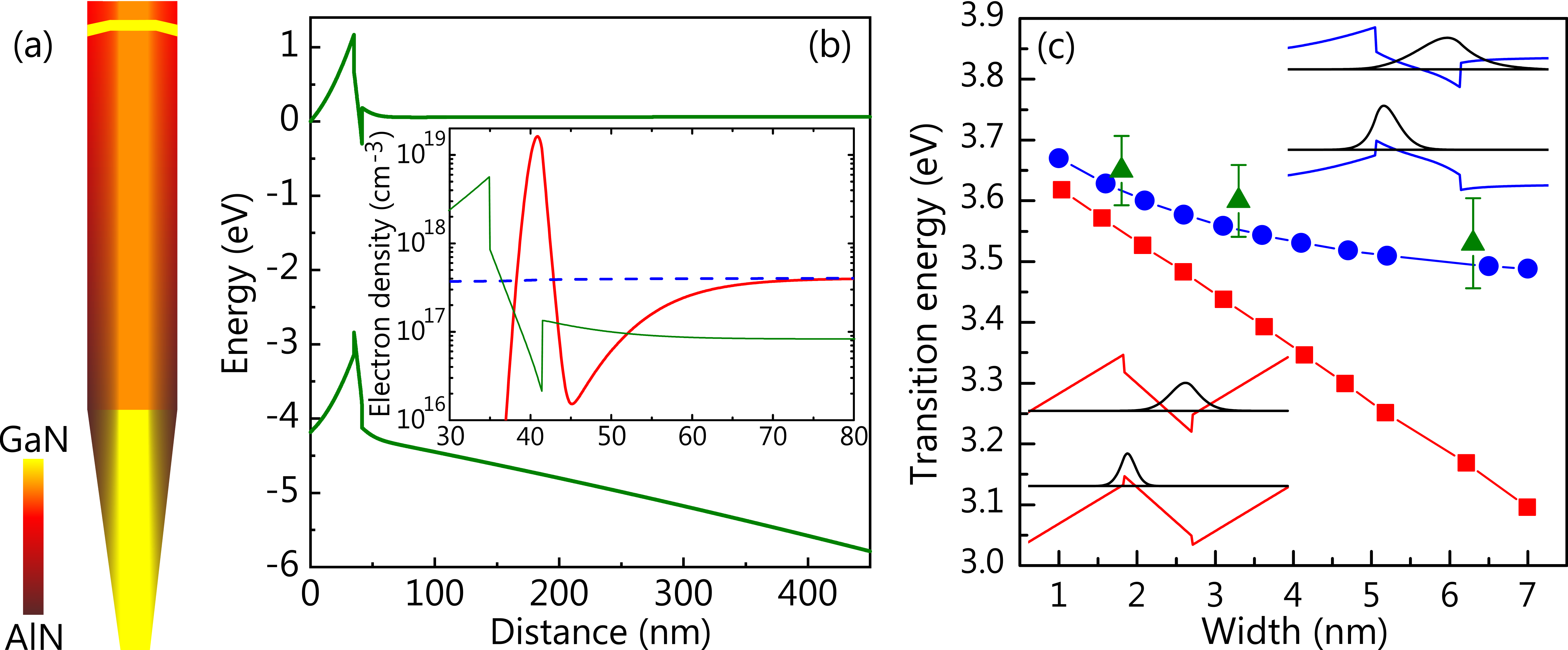}
\caption{\label{fig:simulations} Calculation of the transition energy of the GaN quantum disks with an electron density determined by polarization doping. (a) Schematic representation of the spontaneously formed core/shell structure of the GaN/(Al,Ga)N nanowires under investigation as determined by HAADF-STEM [cf.\ Figs.~\ref{fig:structure}(a)--\ref{fig:structure}(c)], EDX [cf.\ Fig.~\ref{fig:structure}(d)] and APT [cf.\ Figs.~\ref{fig:structure}(e) and \ref{fig:structure}(f)]. The Al content is color-coded according to the scale in the figure. (b) Axial band profile of a planar heterostructure with a 6.5~nm wide GaN quantum well embedded in an (Al,Ga)N layer with a linear grading in the Al content from 0.3 at the surface (0) to 1 at the bottom (450~nm). The inset shows the electron density in this graded (Al,GaN) layer without (blue dashed line) and with an inserted GaN quantum well (red solid line) close to the surface of the structure. A sketch of the conduction band profile is superimposed as thin green line to visualize the position of the GaN quantum well. (c) Transition energy as a function of the quantum well width due to the combined effect of polarization doping and the Al-deficient core with an Al content of 0.1 (circles). For comparison, the transition energy expected without polarization doping is shown by squares. The lines are guides to the eye. The insets show the band profiles and the electron and hole wave functions obtained for these two cases. The experimental peak energies of the PL band of the GaN disks at 10~K are indicated by triangles. The error bars denote the FWHM of the band.}
\end{figure*}

Combining the results of APT with EDX and assuming that the Al content in the core is constant along the length of the (Al,Ga)N segment as suggested by the uniform emission energy observed in CL [Fig.~\ref{fig:PLandCL}(c)], we deduce that the shell essentially consists of pure AlN at the bottom of the (Al,Ga)N segment [i.\,e., at about 400--500~nm in Fig.~\ref{fig:structure}(d)]. Subsequently, the shell composition decreases toward the nanowire top to a value of about 0.4 close to the position of the disk at the top. Figure \ref{fig:simulations}(a) visualizes the resulting complex core/shell structure that forms spontaneously during growth despite a constant temperature and constant fluxes, much like reported previously by \citet{Allah2012} for 300~nm short (Al,Ga)N nanowire segments grown by PAMBE. The drastic deviation from the intended uniform alloy was ascribed to the different diffusion lengths of Ga and Al on the nanowire sidewalls and the geometric shadowing effect occurring during the growth of dense nanowire ensembles \cite{Allah2012}. An axial gradient of the Al content in an (Al,Ga)N segment followed by a constant composition after 200~nm of growth was also reported by \citet{Pierret2013a} In fact, growth would be expected to approach a steady-state regime---leading to a radial composition profile that does not change anymore along the nanowire axis---once the nanowire diameter and density have become constant. In the present case, Fig.~\ref{fig:structure}(d) clearly shows that we do not reach this steady-state conditions despite the considerable length of the (Al,Ga)N segment. The origin of this extended growth under non-stationary conditions is the one order of magnitude lower nanowire density compared to the ensembles studied in Refs.~\citenum{Allah2012} and \citenum{Pierret2013a} (for the density, see Ref.~\citenum{Pierret2013b}), the correspondingly weaker influence of shadowing, and the resulting continuous tapering of the nanowires (see the top- and sideview secondary electron micrographs in the Supporting Information). 

The complex core/shell structure has important consequences for the electronic and optical properties of the nanowires, which have not been elucidated so far. In particular, due to the spontaneous polarization of the group-III nitrides, the compositional grading of the (Al,Ga)N segment along the nanowire axis induces a bulk polarization charge, which in turn leads to the accumulation of mobile charge carriers of the opposite sign. This phenomenon, known as polarization doping \cite{Jena2002}, has been exploited for \emph{n}-type as well as \emph{p}-type doping of GaN/(Al,Ga)N heterostructures \cite{Simon2010}, and is the rationale for the nanowire diode design proposed in Refs.~\citenum{Carnevale2011} and \citenum{Sarwar2015}.

Obviously, the radial and axial compositional gradients as well as the effects of polarization doping are essential for an understanding of the radiative transitions observed in PL and CL [cf.\ Figs.~\ref{fig:PLandCL}(b) and \ref{fig:PLandCL}(d)]. The low energy of the emission band attributed to the (Al,Ga)N segment, and the absence of any signal above 4.1~eV either in the PL [Fig.~\ref{fig:PLandCL}(b)] or CL spectra (not shown here) is a direct consequence of the nanowires' core/shell structure in conjunction with an efficient transfer of photo- or cathodogenerated carriers from the Al-rich shell into the Al-deficient core. In particular, carriers are also transferred from the shell with intermediate Al content despite the existence of a barrier between the shell and the core [cf.\ Figs.~\ref{fig:structure}(e) and \ref{fig:structure}(f)], as demonstrated by the absence of any emission at higher energies. In fact, the radial band profile is affected by two factors that favor transfer of electrons: first, polarization doping resulting in an $n$-type core, and second, Fermi level pinning at the nanowire sidewalls,\cite{Lahnemann2016} both of which induce a band bending resulting in a radial electric field driving electrons toward the core.  

Likewise, electrons induced in the shell by polarization doping transfer to the core and from there to the GaN quantum disk, thus screening its internal electrostatic field. This charge accumulation potentially accounts for the absence of the quantum-confined Stark effect demonstrated by the spectra in Fig.~\ref{fig:PLandCL}(b). However, the conclusion that polarization doping is indeed responsible for this effect would be premature at this stage. In particular, we first have to consider if the unintentional background doping of the nanowires could be high enough to cause this effect as well, and second, we need to quantitatively estimate the electron density induced by polarization doping in comparison.

In general, both GaN layers and GaN nanowires grown by MBE are invariably \emph{n}-type doped due to the unintentional incorporation of O. GaN nanowires synthesized in our lab typically exhibit background doping densities on the order of $10^{16}$~cm$^{-3}$.\cite{Pfuller2010a,Pfuller2010b,Corfdir2014b} This low doping level is also reflected by narrow PL linewidths down to 1~meV and below.\cite{Corfdir2014b} For (Al,Ga)N nanowires, the dominant alloy broadening of the optical transitions impedes a corresponding optical analysis of the background doping. Due to the high reactivity of Al with O, the doping level could very well be substantially higher than in GaN (see, for example, Ref.~\citenum{Elsass2000} for studies of metal-polar films) but we are not aware of data for N-polar (Al,Ga)N. For GaN of either polarity, there is general agreement that O incorporation is reduced substantially with increasing growth temperature.\cite{Koblmuller2007,Cheze2018} Our nanowires were grown at temperatures not accessible for planar layers (820°C), and we would thus expect that O incorporation is reduced even further. For a quantitative estimate, we have examined the APT data for the presence of O, but also C, Ca, and B (which are the most abundant impurities in GaN layers grown by PAMBE). The only element present with a concentration above the detection limit was the isoelectronic impurity B. Due to the unusually high laser power we had to use for evaporation, the APT detection limit may have been higher than under ideal circumstances (on the order of $10^{17}$~cm$^{-3}$)\cite{Koelling2017}, but none of these impurities should be present in concentrations exceeding $10^{18}$~cm$^{-3}$. Note that this concentration of O in (Al,Ga)N does not translate into a correspondingly high electron concentration, since O undergoes a DX transition with increasing Al content.\cite{Gordon2014} It is thus most unlikely that the background doping in our nanowires is sufficiently high to screen the internal electrostatic fields in the GaN quantum disk.  

To quantitatively determine the electron density induced by polarization doping alone and the resulting transition energy, a full three-dimensional simulation of the complex core/shell structure would be desirable. However, when attempting to simulate such a core/shell nanowire including the axial and radial gradients in composition using nextNano\texttrademark, we found that the solution depends sensitively on details of the structure, such as the precise thickness of the shell and the presence or absence of semipolar facets. To illustrate the physical principles, and to get a reliable order-of-magnitude estimate, we thus employ simple one-dimensional simulations \cite{Snider1990,Tan1990} as discussed in the following (for further details, see the \hyperref[sec:methods]{Methods} section).

To estimate the electron density induced by polarization doping, we consider a planar heterostructure with a 6.5-nm-wide GaN well embedded in an (Al,Ga)N layer with a negative Al gradient along the growth axis (the $[000\bar{1}]$ direction). Figure \ref{fig:simulations}(b) shows the band profile of this heterostructure where the Al content is assumed to vary linearly from 1 to 0.35 from the bottom to the top of the structure, similar to the variation of the Al content in the Al-rich shell of the top segment of our nanowires containing the GaN quantum disk [cf.\ Fig.~\ref{fig:structure}(d)]. The negative gradient of the Al content along the $[000\bar{1}]$ direction leaves a three-dimensional slab of positive fixed charge, which is compensated by free electrons from both shallow donors in the bulk and/or surface donor states. This polarization-induced doping manifests itself by the change of band gap occurring entirely by a variation of the valence band edge, while the conduction band edge remains close to the Fermi level. 

The dashed line in the inset of Fig.~\ref{fig:simulations}(b) shows the polarization-induced free electron density in a graded (Al,Ga)N segment without GaN well to be on the order of $5 \times 10^{17}$~cm$^{-3}$. The solid line in the inset shows the electron concentration in a structure with a graded (Al,Ga)N segment and additional GaN quantum well, corresponding to the sample whose band profile is shown in the main figure. The charge transfer taking place in this structure results in a degenerately doped GaN well with an electron sheet density $n_s$ of $4\times10^{12}~\mathrm{cm^{-2}}$. 

To quantitatively determine the impact of this high electron density on the transition energies in the GaN quantum well, we next calculate the electron and hole states in GaN quantum wells with a width between 1 and 7~nm embedded in Al$_{0.1}$Ga$_{0.9}$N, the approximate composition of the nanowire core. The calculations are done for undoped wells, for which the polarization results in an internal electrostatic field of around 1~MV/cm,\cite{Dreyer2016} and for wells with an electron density as determined by the simulations discussed above. We also take into account the compressive strain in the GaN quantum well imposed by a coherent interface with the Al$_{0.1}$Ga$_{0.9}$N layer. Figure~\ref{fig:simulations}(c) shows the transition energies for these two cases together with the experimental peak energies of the low-energy PL bands obtained by PL spectroscopy at 10~K (not shown here). The transition energies for the undoped wells are governed by the quantum confined Stark effect, i.\,e., they depend linearly on width with a slope given by the internal electrostatic field. As expected, they fall below the GaN band gap (3.504~eV) for a well width exceeding 2.5~nm. In contrast, the quantum wells with an electron density equal to that induced in the disks by polarization doping in the shell exhibit a transition energy above the GaN band gap even for the widest well, in close agreement with the experimental data. This result shows conclusively that the primary reason for the absence of the quantum confined Stark effect is polarization doping.

The insets of Fig.~\ref{fig:simulations}(c) illustrate that the screening of the internal fields increases the electron-hole overlap in the quantum disk. Whereas electrons and holes are spatially largely separated in the undoped disk (bottom left), they have an appreciable overlap for the disks with a high electron sheet density (right top). This enhanced electron-hole overlap is expected to affect the recombination dynamics of the electron-hole pairs within the GaN disk and consequently the internal quantum efficiency. Even more importantly, carrier recombination in these structures is affected not only quantitatively, but also qualitatively: the electron density in the disk is well above the Mott density,\cite{Binet1999,Rossbach2014a} and recombination is thus no longer excitonic, but occurs between the two-dimensional electron gas and photoexcited holes.

To investigate the consequences of this high electron density for the recombination dynamics in the GaN disk, we perform time-resolved PL measurements on the nanowire ensemble with the 6.3~nm-thick GaN disk at temperatures between 5 and 300~K. The disks were excited directly using an excitation wavelength of 325~nm, corresponding to an energy below the bandgap of the (Al,Ga)N core. The PL transients presented below are thus not affected by carrier transfer processes.
\begin{figure*}[h!]
	\includegraphics[width=1\textwidth]{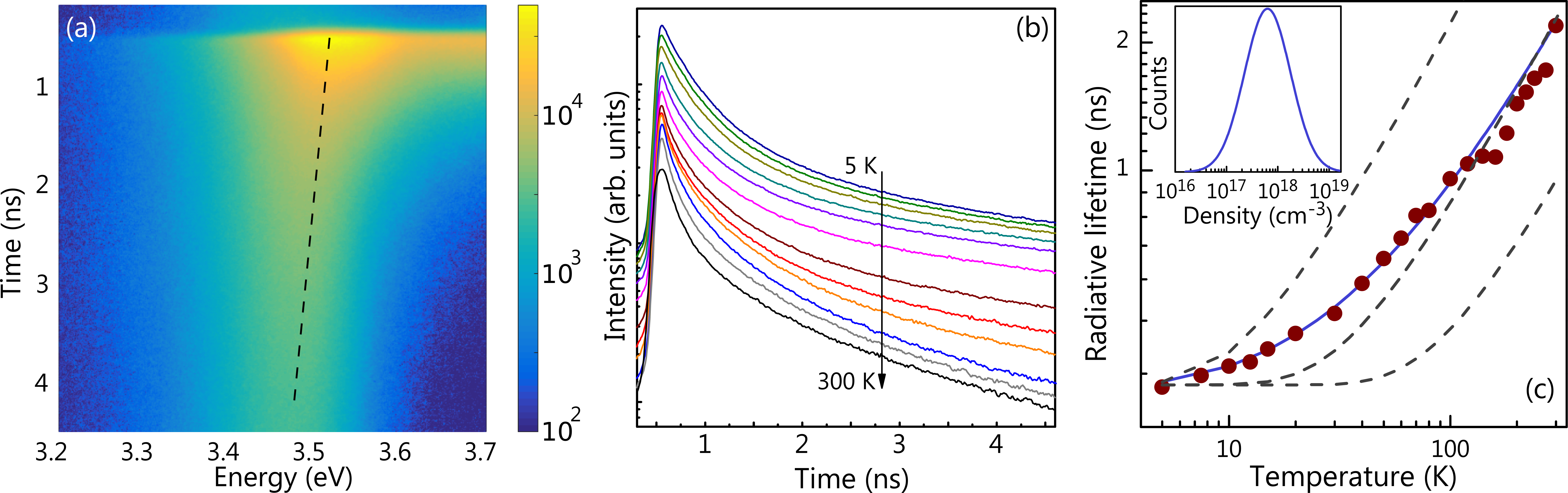}
	\caption{\label{fig:recombDynamics} Recombination dynamics of the nanowire ensemble with a 6.3~nm thick GaN disk. (a) Streak camera image obtained upon pulsed excitation at 325~nm and 10~K. The dashed line highlights the spectral diffusion of the emission band from the GaN quantum disk. (b) Spectrally integrated PL intensity transients obtained from streak camera images acquired at temperatures between 5 and 300~K (from top to bottom: 5, 10, 20, 40, 50, 70, 100, 120, 140, 180, 220 and 300~K). (c) Temperature dependence of the radiative lifetime determined from the peak intensity of the transients (solid circles). The dashed lines show the calculated temperature dependence for electron sheet densities in the quantum disk of $0.2$, $0.6$, and $1.6\times10^{12}~\mathrm{cm^{-2}}$ (from left to right). The solid line represents the radiative lifetime of a nanowire ensemble with an ensemble distribution of the electron volume density as shown in the inset.}
\end{figure*}

Figure~\ref{fig:recombDynamics}(a) shows a representative streak camera image at 10~K. The emission band of the GaN quantum disk is observed to redshift by 30~meV during the initial 4~ns of the PL decay. For polar GaN/(Al,Ga)N quantum wells, such a temporal redshift may occur as a consequence of a dynamical screening of the internal electrostatic fields induced by a high excitation density.\cite{Reale2003,Lefebvre2004} In the present case, this phenomenon can be safely ruled out, since the photogenerated carrier density $\Delta n$ inside the disk is at most $2\times10^{10}\mathrm{cm^{-2}}$, i.\,e., more than two orders of magnitude smaller than the electron sheet density resulting from polarization doping ($n_{s} \approx 4\times10^{12}~\mathrm{cm^{-2}}$).

Another possible reason for this gradual redshift is the progressive relaxation of photoexcited excitons within a band of localized states, leading to their  spectral diffusion.\cite{Holstein1977,Gobel1982} This phenomenon is most frequently observed in ternary layers and quantum wells, for which compositional fluctuations induce the disorder responsible for exciton localization.\cite{Cohen1982,Donegan1994,Sun2003} In the present case, compositional fluctuations in the barriers or steps at the GaN/(Al,Ga)N interfaces could potentially induce localization as well.\cite{Gallart2000} However, the transfer of carriers from higher to lower energy states manifests itself in a characteristic temporal delay of the emission at lower energies, which is not observed here. Instead, the rise time of the emission is found to be the same regardless of energy. We also do not observe any other signature of localization effects, such as the typical \textsf{S} shape of the transition energy with temperature (not shown).\cite{Li2005,Kuokstis2006,Grundmann2009}

An alternative explanation for the spectral shift suggests itself when considering that we are not dealing with a spatially uniform emitter, but a large  ensemble of individual emitters (in the present experiments, we excite about $10^4$ nanowires). In other words, the temporal redshift may simply result from a spectral superposition of GaN quantum disks emitting at different energies. In fact, for the present nanowire ensembles with a spatially random arrangement, it is not conceivable that the Al gradient in the shell is identical for each nanowire. Differences must occur because of both shadowing effects \cite{Allah2012,Pierret2013b} and the different diameters of individual nanowires,\cite{Brubaker2019} affecting the diffusion of the group-III adatoms on the nanowire sidewalls. These differences in the Al gradient in turn result in different electron densities induced by polarization doping. The higher the electron density in the quantum disk, the lower the internal electrostatic field due to screening, which translates into higher emission energies and shorter lifetimes. The most heavily doped quantum disks thus dominate the high-energy emission at short times. In more lightly doped disks, the residual field leads to longer lifetimes and lower emission energies, which lead to the long-living and redshifted tail of the emission. The net result of this superposition of an ensemble of disks with different transition energies and lifetimes is the apparent spectral shift of the emission band as seen in Fig.~\ref{fig:recombDynamics}(a).

Figure~\ref{fig:recombDynamics}(b) shows PL transients spectrally integrated over the entire emission band between 5 and 300~K. The transients are characterized by a continuously changing slope on a semilogarithmic scale and can be adequately described only by the sum of at least three exponentials. This finding is in full agreement with the hypothesis discussed above: the spectrally integrated emission is a superposition of the individual (exponential) transients of nanowires with different electron densities, and thus different lifetimes.  Approximating the fast and slow components of the transient at 5~K by single exponentials, we obtain decay times of about 0.4 and 4~ns, respectively. For comparison, planar GaN/(Al,Ga)N quantum wells of similar width and average doping density were found to exhibit a lifetime of 2~ns.\cite{Thamm2000} The maximum electron density in our nanowire ensemble is thus significantly higher than the average value of $n_s$ determined from the average transition energy. 

For temperatures above 20~K, the decay is seen to accelerate accompanied by a continuously decreasing integrated intensity [cf.\ Fig.\ \ref{fig:recombDynamics}(b)], revealing an increasing participation of nonradiative decay channels (see also the comparison of resonant and nonresonant excitation in the Supporting Information). To separate radiative and nonradiative channels, we examine the peak intensity of the transients, which is directly proportional to the radiative decay rate.\cite{Sermage1989,Deveaud1991,Brandt1996} Figure~\ref{fig:recombDynamics}(c) shows the temperature dependence of the radiative lifetime determined in this way. The radiative lifetime starts to increase already at temperatures as low as 10~K, confirming that the emission from the GaN quantum disks is dominated by the radiative recombination of delocalized charge carriers already at low temperatures. However, the increase in radiative lifetime does not enter the linear temperature dependence characteristic for nondegenerate two-dimensional systems in general and GaN/(Al,Ga)N quantum wells in particular.\cite{Lefebvre1998,Rosales2013} To quantitatively understand this peculiar temperature dependence, we calculate the spontaneous recombination rate using Fermi's golden rule for the general case of Fermi-Dirac statistics valid also for the high carrier density in the GaN quantum disks (for details, see the Supporting information) \cite{Arakawa1984,Christen1989,Oliveira1993}. Note that this calculation ignores any residual electric fields in the disk. Consequently, we do not compare the absolute transition rate in the following, but only its temperature dependence. 

The three dashed lines in Fig.~\ref{fig:recombDynamics}(c)  show the inverse radiative rates obtained by our calculations for electron sheet densities of $n_{s}=0.2$, $0.6$, and $1.6\times10^{12}~\mathrm{cm^{-2}}$. In contrast to the strictly linear dependence of a nondegenerate quantum well, the radiative rate for high electron sheet densities is constant at low temperatures, for which the system becomes degenerate. The higher the electron density, the higher the temperature up to which degeneracy prevails, and thus the larger the temperature range of constant radiative lifetime (for more details, see the Supporting Information).    

The theoretical dependence for an electron density of $n_{s}=6\times10^{11}~\mathrm{cm^{-2}}$ is closest to the experimental data, but it does not describe them well, particularly at low temperatures. The agreement can be much improved when allowing for a distribution of the electron density as qualitatively discussed above in the context of Figs.~\ref{fig:recombDynamics}(a) and \ref{fig:recombDynamics}(b). Quantitatively, we obtain the solid line in Fig.~\ref{fig:recombDynamics}(c) by assuming a log-normal distribution of $n_s$ with a mean of $2\times10^{12}~\mathrm{cm^{-2}}$ or a volume density of $3\times10^{18}~\mathrm{cm^{-3}}$ as shown in the inset of Fig.~\ref{fig:recombDynamics}(c), not too far from the average value estimated above as a result of polarization doping ($n_{s}=4\times10^{12}~\mathrm{cm^{-2}}$). This agreement suggests that despite the crudeness of our approach concerning the energy and the lifetime of carriers in the GaN quantum disks in our nanowires, we have reached a fairly coherent understanding of both the static and dynamical electronic properties of these disks. In particular, the modest thermal quenching of the emission from the GaN quantum disks (see Supporting Information) is understood to be a result of the short radiative lifetime due to the high electron density and not a particularly high material quality.

To summarize and conclude, we have shown that, during the synthesis of GaN/(Al,Ga)N nanowire ensembles by molecular beam epitaxy, a complex core/shell structure forms spontaneously, characterized by a continuously varying Al composition both along the growth axis and across the diameter of the nanowires. This phenomenon has been observed in similar form by other groups, which shows that it is not caused by specific growth conditions, but the fundamental growth mechanisms of (Al,Ga)N nanowires by PAMBE. The peculiar structure resulting from this self-assembly process has important consequences for the optoelectronic properties of the nanowires. First of all, the axial Al gradient in the Al-rich shell induces a three-dimensional polarization charge, leading to the accumulation of electrons in the shell. Second, due to the Al deficient core, these electrons in the shell are efficiently transferred to the GaN quantum disk. The resulting high electron sheet density inside the GaN disk screens the internal electrostatic field, effectively canceling the quantum-confined Stark effect, and manifesting itself in both transition energies almost independent of thickness and short radiative lifetimes up to room temperature that result in a comparatively high internal quantum efficiency. The latter is obviously attractive for the realization of efficient light emitting devices in the ultraviolet range. However, we emphasize that for taking advantage of the effects reported here in any application, one would have to achieve a high level of control over the formation of the compositional gradients. To reach this aim, further studies are required to fully understand the mechanisms leading to the self-assembly of (Al,Ga)N nanowires in PAMBE. Moreover, this compositional self-assembly is likely to occur also for other mixed cation nanowires synthesized by diffusion-controlled growth techniques such as MBE. Hence, our findings may be relevant not only for deep UV emitters based on (Al,Ga)N nanowires, but for example also for  MBE-grown (In,Ga)N nanowires for emission in the visible spectrum and solar energy harvesting.

\section{Methods}
\label{sec:methods}

\subsection{Nanowire synthesis}
The GaN/(Al,Ga)N nanowire ensembles are synthesized utilizing a DCA Instruments P600 molecular beam epitaxy system equipped with two radio frequency plasma sources for active N and solid-source effusion cells for Al and Ga. First, GaN nanowires are grown on Si$(111)$ substrates relying on their spontaneous formation on this substrate under suitable experimental conditions.\cite{Garrido2013,Garrido2015}. The growth temperature is measured with an optical pyrometer calibrated with the $(1 \times 1) \rightarrow (7 \times 7)$ surface reconstruction transition of Si$(111)$, which takes place at 860°C. Fluxes calibrated using GaN and AlN films are given in units of monolayers (ML)/s where 1~ML refers to the cation density of $1.136 \times 10^{15}$~cm$^{-2}$ on GaN$(000\bar{1})$. A two-step growth approach is employed in order to minimize the incubation time.\cite{Zettler2015} Specifically, the substrates are outgassed at 885°C for 30~min prior to growth to remove any residual Si$_x$O$_y$ from the surface. Afterward, the N shutter is opened, and the substrates are exposed for 10~min to an active N flux of 0.74~ML/s at 790°C.  Nanowire nucleation is initiated by opening the Ga shutter providing a Ga flux of 0.34~ML/s. Once nanowire nucleation has commenced as observed by reflection high-energy electron diffraction, the temperature is increased to 820°C, and growth of the GaN nanowire base is continued. After the formation of a GaN nanowire ensemble with a nominal length of 750~nm, the Ga flux is reduced to 0.24~ML/s and an Al flux of 0.1~ML/s is added for the synthesis of a nominally 800~nm long (Al,Ga)N segment with a nominal Al content of 0.3 and the same V/III ratio as for the GaN nanowires underneath. For the formation of the GaN quantum disk, the Al shutter is simply closed for a duration corresponding to the intended disk thickness. Finally, the disk is capped by a nominally 35~nm long (Al,Ga)N segment. Three samples are grown in this way, differing only in the thickness of the GaN quantum disk with nominal values of 1.8, 3.3, and  6.3~nm. A sample without quantum disk, but otherwise identical (Al,Ga)N nanowires on a GaN base serves as comparison. The nanowires thus obtained have an average length of 1.55~\textmu m (close to the expected value) and an average equivalent disk diameter at their top of 125~nm. The mean density of nanowires in the ensemble is $3 \times 10^{9}$~cm$^{-2}$. Top- and sideview secondary electron micrographs of the sample with a 6.3-nm-thick-disk are shown in the Supporting Information.

\subsection{Morphology and microstructure}
The morphological properties of the as-grown GaN/(Al,Ga)N nanowires are studied by scanning electron microscopy carried out in a Hitachi S-4800 field emission microscope using an acceleration voltage of 5~kV (see Supporting Information). Single nanowires in cross-sectional specimens are investigated by transmission electron microscopy. The cross-sections are prepared by mechanical grinding and dimpling followed by Ar-ion beam milling down to electron transparency.  Scanning transmission electron microscopy is performed by using a JEOL 2100F field emission instrument equipped with a bright-field and a dark-field detector and operated at 200~kV. High-angle annular dark-field micrographs of the nanowires reflect the chemical contrast between the GaN quantum disk and the surrounding (Al,Ga)N segments.

\subsection{Composition}
The composition of single nanowires is investigated by energy dispersive x-ray spectroscopy and atom probe tomography. The former is performed using an EDAX Apollo XV silicon drift detector attached to a Zeiss Ultra55 field emission scanning electron microscope operated at 5~kV. The latter is carried out using a Cameca LEAP 4000X-HR system. Single nanowires from the ensemble with the 6.3~nm thick quantum disks are isolated in an FEI Nova Nanolab 600i. Welds are created by electron-induced metal deposition of Pt or Co. The selected nanowire is mounted on a tip, and the evaporation of atoms is triggered by a laser generating picosecond pulses at a wavelength of 355~nm. Since the (Al,Ga)N nanowire absorbs only weakly at this wavelength, a comparatively high pulse energy of 100~pJ is used. Due to the increase in Al content toward the bottom of the nanowire, only the top 100~nm of the nanowire are evaporated and analyzed. The field-of-view probed in these experiments is about 40~nm.

\subsection{Spectroscopy}
For the investigation of the optical properties of the nanowire ensembles, the as-grown samples are mounted onto the cold finger of a liquid He cryostat allowing continuous control of the sample temperature between 10 and 300~K. Cathodoluminescence spectroscopy is carried out at room temperature with a Gatan Mo\-noCL4 system fitted to a Zeiss Ultra55 field-emission scanning electron microscope operated at 5~kV with a probe current of 0.7~nA. The signal is spectrally dispersed by a monochromator and detected using a photomultiplier tube. Continuous-wave photoluminescence spectroscopy is performed by exciting the ensembles with the 244~nm (5.081~eV) line of a Coherent Innova 90C FreD Ar$^+$ ion laser focused to a spot of about 0.5\,\textmu m diameter yielding an excitation density on the order of 1~W/cm\(^2\). The signal is spectrally dispersed by a monochromator and detected with a cooled charge-coupled device array. Time-resolved PL measurements are performed by exciting the ensemble with the second harmonic (325~nm) of 200-fs-pulses from an APE optical parametric oscillator synchronously pumped by a Coherent Mira 900 Ti:sapphire laser, which itself is pumped by a Coherent Verdi V10 frequency-doubled Nd:YVO$_{4}$ laser. The pulses with an energy of 4~pJ are focused onto the sample with a 75-mm plano-convex lens to a spot with a diameter of 30\,\textmu m. The transient PL signal is dispersed by a monochromator and detected by a Hamamatsu C5680 streak camera with a temporal resolution set to about 50~ps.

\subsection{Simulations}
The band profiles and electron densities are obtained by one-dimensional Schrödinger-Poisson simulations performed with $1$DPoisson.\cite{Tan1990} For the polarization of Al$_x$Ga$_{1-x}$N, we assume a value of $-2 \times 10^{-6}\, (1+2.1 x + 0.475 x^2)$~C/cm$^2$ (the negative sign is due to the fact that the nanowires grow in the $[000\bar{1}]$ direction)\cite{Garrido2013}. This value is about half of that predicted theoretically, but in agreement with the majority of experimental results.\cite{Dreyer2016} For the conduction band offset between GaN and (Al,Ga)N, we take the value reported by \citet{Tchernycheva2006} 
The spontaneous recombination rate is calculated from Fermi's golden rule assuming $k$ conservation as $\mathcal{R}_{\text{sp}}=R_{0}\int_{0}^{\infty}\rho_{r}(E)f_{c}(E)[1-f_{v}(E)]dE$, where $\rho_{r}(E)$ is the reduced density of states, $f_{c}(E)$ and $f_{v}(E)$ are the Fermi-Dirac distributions for the conduction and the valence bands, respectively, and $R_{0}$ is a prefactor. Details of this calculation are presented in the Supporting Information. 

\begin{suppinfo}
	Top- and sideview secondary electron micrographs of the sample with a 6.3-nm-thick disk, secondary electron micrograph of the nanowire specimen prepared for APT, temperature dependence of the integrated photoluminescence intensity for excitation wavelengths of 325 and 244~nm, calculation of the spontaneous recombination rate. 
\end{suppinfo}

\begin{acknowledgement}
	The authors thank Jesús Herranz for a critical reading of the manuscript. P.\,C. is grateful to the Fonds National Suisse de la Recherche Scientifique for funding through project 161032. J.\,K.\,Z. and T.\,A. acknowledge the financial support received by Deutsche 	Forschungsgemeinschaft within SFB951 and by Bundesministerium für Bildung und Forschung through Project No.\ 	FKZ:13N13662, respectively. S.\,F.\,G. acknowledges the partial financial support received through the Spanish program Ramón y Cajal (cofinanced by the European Social Fund) under Grant No. RYC-2016-19509 from Ministerio de Ciencia, Innovación y Universidades. 
\end{acknowledgement}

\bibliography{bibliography}

\end{document}